\documentclass[aip,apl,amsmath,amssymb,preprint,reprint]{revtex4-1}
\usepackage[dvipsnames]{xcolor}
\usepackage{siunitx}
\usepackage{graphicx}
\usepackage{dcolumn}
\usepackage{bm}

\usepackage[utf8]{inputenc}
\usepackage[T1]{fontenc}
\usepackage{mathptmx}
\usepackage{etoolbox}
\usepackage[normalem]{ulem}

\newcommand{\om}[1]{\omega_\textrm{#1}}
\newcommand{\Om}[1]{\Omega_\textrm{#1}}
\newcommand{\ke}{\kappa_\textrm{ext}}
\newcommand{\gammam}{\gamma_\textrm{m}}

\newcommand{\Pd}{P_\textrm{d}}

\makeatletter
\def\@email#1#2{%
 \endgroup
 \patchcmd{\titleblock@produce}
  {\frontmatter@RRAPformat}
  {\frontmatter@RRAPformat{\produce@RRAP{*#1\href{mailto:#2}{#2}}}\frontmatter@RRAPformat}
  {}{}
}%
\makeatother
\begin{document}

\preprint{AIP/123-QED}

\title{Cryogenic Magnomechanics for Thermometry Applications}

\def\toeditor#1{\textcolor{Maroon}{\textit{#1}}}
\def\nrcaddr{National Research Council Canada, Metrology Research Centre, 1200 Montreal Road, Ottawa, Ontario K1A 0R6, Canada.}
\def\uabaddr{Department of Physics,
University of Alberta,
4-181 CCIS,
Edmonton, Alberta T6G 2E1, Canada }
\author{Y. Huang}\email{yunhu@ualberta.ca}
\affiliation{\uabaddr}
\author{P.M.C.\ Rourke}
\author{A.\ Peruzzi}
\author{J.\ Jin}
\affiliation{\nrcaddr}
\author{M.\ Ebrahimi}
\author{A.\ Rashedi}
\author{J.\ P.\ Davis}%
\affiliation{\uabaddr}

\date{\today}

\begin{abstract}
Cavity magnomechanics combines strong coupling between magnons in a dielectric material and microwave cavity photons with long-lived mechanical resonances.  Forming a triple resonance condition, this hybrid quantum system promises many advantages in quantum technologies, yet has never been studied at the cryogenic temperatures required to reveal such quantum properties.  We report the observation of magnomechanics at cryogenic temperatures down to \qty9K. The experiment was conducted using a YIG sphere inside a microwave cavity, where we measured both the thermomechanical motion and the temperature-dependence  of the magnon linewidth.
\end{abstract}

\maketitle
In the efforts underway to develop useful quantum technologies, hybrid quantum systems are expected to play a significant role.  In particular, by combining the best features of two or more disparate systems, hybrid systems hold the promise for advancing such quantum subsystems as quantum memories \cite{Groblacher_quantum_memory,LeBlanc}, quantum wavelength transducers \cite{Groblacher,Safavi-Naeini,Regal2022}, and nonreciprocal devices such as circulators and directional amplifiers \cite{Metelmann,Teufel_nonreciprocal}.  One of the most recent hybrid quantum systems to gain interest is that of cavity magnonics \cite{Huebl,Zhang2014, Tabuchi2014, Tobar, Rameshti, Lachance-Quirion_2019, Gloppe2019}, along with the related area of cavity magnomechanics.  Cavity magnonics is notable for the relative ease with which strong coupling, or even ultrastrong coupling \cite{Potts_tuneable,Tobar,Bourcin}, can be achieved between magnons and cavity microwave photons \cite{Flatte}.  Using cavity magnonics exciting physics has been observed such as coherent coupling to a qubit \cite{Tabuchi} and unidirectional invisibility \cite{Hu_uni_invis}. 

Including the mechanical mode of the magnonic object opens up even more possibilities.  Importantly, the mechanical mode frequency can be tuned such that it is aligned with the splitting between the hybrid cavity polariton modes, forming a triple resonance system that cavity enhances both pump and probe Raman tones \cite{XufengZhang2016, Potts2021}.  Following the discovery of cavity magnomechanics \cite{XufengZhang2016}, there have been a plethora of theoretical proposals for novel quantum applications of this triply resonant system, including squeezing \cite{Li_mwsqueezing}, wavelength transduction \cite{Sylvia}, and long-range entanglement \cite{Sohail2023}.  Despite this, very few experimental demonstrations of cavity magnomechanics have been realized \cite{XufengZhang2016,Potts2021,KerrNonlinearity} and none at the cryogenic temperatures needed to realize the quantum properties of this system.  Here, we demonstrate the cryogenic cavity magnomechanics, showing thermalization of the mechanical mode down to $\qty9K$.  This is aided by additional measurements of the magnon linewidth as a function of temperature, which are used as a secondary thermometer, confirming thermalization of the hybrid system.

\begin{figure}[tbh]
\includegraphics[width=0.9\columnwidth]{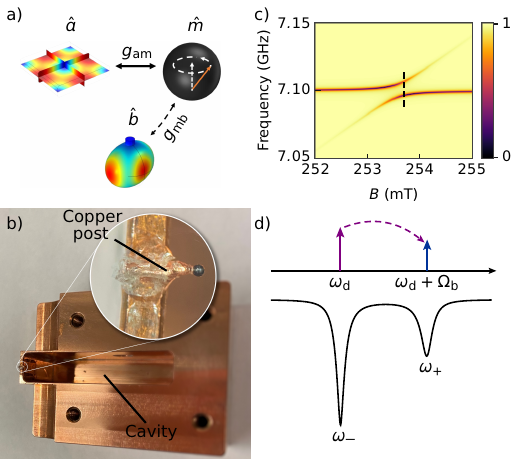}
\caption{\label{fig:cavity}The experimental setup. a) A schematic drawing illustrating the interactions between subsystems: cavity photons ($\hat a$) are coupled to magnons ($\hat m$) with coupling constant $g_\textrm{am}$, and the magnons are coupled to phonons ($\hat b$) with coupling constant $g_\textrm{mb}$. The dashed arrows for $g_\textrm{mb}$ denotes the fact that it is much weaker than $g_\textrm{am}$.  b) The microwave cavity used for the measurements. The circular insert provides a magnified view of the YIG sphere mounted on a copper post. c) Normalized reflection spectra $|S_{11}|$ of the cavity with YIG sphere as a function of the bias magnetic field $B$. Note the avoided level crossing as a result of the strong coupling and hybridization between photons and magnons. d) Slice of data taken from c) at the dashed line ($B=\qty{253.7}{mT}$), illustrating the triple resonance regime. The bottom is the $S_{11}$ reflection spectrum of the microwave cavity with the YIG sphere, given by Eq.~\ref{eq:spectrum}, where hybridized peaks at frequencies $\omega_-$ and $\omega_+$ are tuned to satisfy $\omega_+ - \omega_- = \Omega_\textrm{b}$ (see text, Eq.~\ref{eq:peak_separation}); the top portion shows a probe tone $\omega_\textrm{d}$ that coincides with $\omega_-$, whose blue sideband $\omega_\textrm{d} + \Omega_\textrm{b}$ is then at $\omega_+$, satisfying the triple resonance condition. There is also a red sideband at $\omega_\textrm{d} - \Omega_\textrm{b}$, but it is not cavity enhanced and therefore its amplitude is negligible compared to the blue sideband.}
\end{figure}

Measurements were performed using a yttrium iron garnet (YIG) sphere, where the mechanisms of magnonics and magnomechanics have been described in detail previously\cite{XufengZhang2016,Potts2021}.  In brief, the large number of spins in the YIG sphere can act as a single magnon resonance, called the Kittel mode.  This mode is excited by, and coupled to, the magnetic field of a 3D microwave cavity, as shown in Fig.~1. 
The frequency of the Kittel mode, $\om m$, is determined by $\om m = \gamma|\mathbf{B_0}|$, where $\mathbf{B_0}$ is the bias magnetic field, and $\gamma$ the gyromagnetic ratio $\gamma/2\pi = \qty{28}{GHz/T}$. As the resonance frequency of the Kittel mode is tuned towards the resonance of the cavity by application of an external magnetic field, the characteristic avoided level crossing occurs, signifying strong coupling and hybridization.  This magnon mode hybridizes with the resonant cavity mode of frequency $\om a$, resulting in polariton modes whose reflection spectrum is given by\cite{XufengZhang2016}
\begin{equation}\label{eq:spectrum}
    S_{11}(\omega) = 1 - \frac{\mathrm{e}^{i\eta_\textrm{mis}}\ke}{i(\omega - \om a) + \frac{\kappa_\textrm{int} + \kappa_\textrm{ext}}2 + \frac{g_\textrm{am}^2}{i(\omega - \om m) + \gamma_\text{m}/2} },
\end{equation}
where $\eta_\textrm{mis}$ is the impedance mismatch, $\kappa_\textrm{ext}$ the external coupling rate, $\kappa_\textrm{int}$ the intrinsic loss rate of the cavity, $\gamma_\textrm{m}$ the magnon loss rate, and $g_\textrm{am}$ the magnon-photon coupling strength. This spectrum has two minima at frequencies $\omega_+$ and $\omega_-$:
\begin{equation}\label{eq:peak_separation}
    \omega_\pm = \frac{\om a + \om m}2 \pm
            \frac{\sqrt{4g_\textrm{am}^2 + (\om a - \om m)^2}}2.
\end{equation}
Note that the frequency spacing between the two polariton modes can be tuned by changing the bias field $\mathbf{B_0}$, which in turn changes $\om m$.  It is noteworthy that the splitting between the polariton modes can be arranged, by the geometry of the microwave cavity and the applied magnetic field, such that it is equal to a mechanical resonance frequency in the YIG sphere, $\Omega_\textrm{b}$.

When a probe tone of frequency $\om{d}$ is sent into the microwave cavity, it interacts with the vibrational phonons of the YIG sphere with frequency $\Om b$ and produces two sidebands at $\om d\pm \Om b$. If either sideband is near a minimum of the polariton spectrum, it becomes cavity enhanced, resulting in that sideband peak becoming amplified. A special case of this amplification occurs if the probe frequency is also near the other polariton spectrum minimum, in which case both the probe frequency and a sideband become cavity enhanced, making the amplification much larger. This triple resonance condition (see Fig.~\ref{fig:cavity}d) is a powerful tool, for example allowing us to observe small thermomechanical signals in the present experiment.

The key part of our experimental system consists of a $\qty{250}{\um}$-diameter YIG sphere inside a 3D microwave cavity. The cavity is made of polished OFHC copper and has an inner dimension of $\qty{30}{mm}\times\qty{30}{mm}\times\qty{0.6}{mm}$, and a TE$_{101}$ mode frequency of $\omega_\textrm{a} = \qty{7.074}{GHz}$, similar to our previous experiments\cite{Potts2021, Potts2023}. In those previous experiments, we placed the YIG sphere in a glass capillary so that the sphere could move freely,  minimizing damping losses. For the current low temperature experiments, however, we determined in preliminary tests that such a mounting method prevented the YIG sphere from sufficiently thermalizing when a drive microwave field is applied, causing the sphere to be significantly hotter than the cryogenic bath. Thus, we instead opted to affix the sphere to the tip of a copper needle (made by sharpening a copper wire), which in turn is attached to the wall of the cavity, both with UV-cured optical adhesive.  Compared to mounting via capillary, this method improved thermalization of the sphere, but at the cost of increased mechanical loss due to clamping: the mechanical decay rate increased from $\sim \qty{100}{Hz}$ with the capillary\cite{Potts2021} to $\sim\qty{1}{kHz}$ with copper needle.

\begin{figure}
\includegraphics[width=0.9\columnwidth]{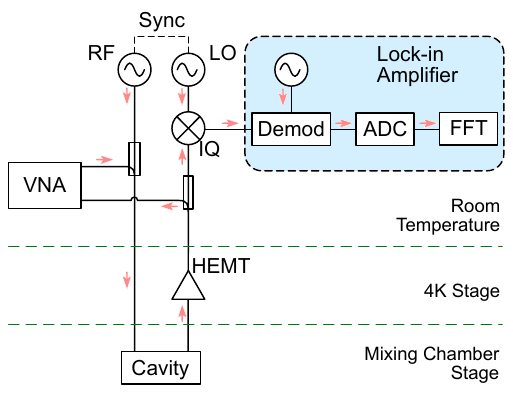}
\caption{\label{fig:setup}%
Schematic of the experimental setup. The microwave signals from the RF source ($\om d$) and the VNA were sent to the cavity, and the reflected signals were amplified by a high-electron-mobility transistor (HEMT) amplifier before being transmitted up to the room temperature stage. When tuning the polariton spectrum or locating the mechanical modes using magnomechanically induced transparency (MMIT), the VNA was turned on to sweep and measure the targeted frequency range. When taking homodyne measurements, the VNA was switched off to avoid undesirable beat frequencies. The reflected RF frequency and its sidebands were sent to an IQ-mixer whose local oscillator (LO) also runs at frequency $\om d$, effectively down-mixing the spectrum so the center frequency is now at DC. The output of the mixer was then fed to the lock-in, demodulated by a frequency close to the mechanical frequency, the resulting time-domain data digitized by its analog/digital converter (ADC), and fast-Fourier transformed (FFT) to eventually produce a frequency spectrum centered around the mechanical frequency.}
\end{figure}

The cavity was inserted into a magnetic yoke structure that contains both permanent magnets and a solenoid, similar to that used in Ref.~[\onlinecite{Tabuchi}]. The permanent neodymium magnets provided the vast majority of the $\sim\qty{0.2}{T}$ bias magnetic field needed to saturate the YIG magnetization, while the solenoid, wrapped around a pure iron core, provided the remaining field -- up to $\sim\pm\qty{50}{mT}$ -- to facilitate sufficient tunability of the magnon frequency.  The cavity and magnet were attached to a gold-plated copper frame, and this assembly was mounted to the low-temperature plate of a cryostat.  Microwave drive tones were applied to the cavity through attenuated coax, and output signals were returned via NbTi coax followed by amplification using a HEMT amplifier at $\qty4K$ before being transmitted to room temperature, Fig.~\ref{fig:setup}.

The microwave signals are coupled to the cavity by way of a coupling pin.  During preliminary runs, we determined that the magnon linewidth $\gammam$ increases significantly in certain low temperature ranges (described further below). In order to maximize the output signal, we chose to set the external coupling $\ke$ to $\sim\qty{8}{MHz}$ so that when the cavity is cooled towards $\qty4K$ it is close to critically coupled. These values for $\ke$ were substantially higher than what we normally used at room temperature, causing stronger extraction of signals from the cavity at the cost of higher loss rate.

Two different schemes were employed in measuring the magnomechanics. The first one used a vector network analyzer (VNA) to measure magnomechanically induced transparency/absorption (MMIT/MMIA)\cite{XufengZhang2016}. Microwave photons of frequency $\omega_\textrm{d}$ are injected into the cavity, some of which up- or down-converted by one mechanical frequency $\Omega_\textrm{b}$. These converted photons coherently interfere with the swept VNA tone and produce transparency or absorption windows as a result, analogous to optomechanically induced transparency (OMIT).\cite{CavityOptomechanics} The mechanics is driven by input photons via hybridized photon-magnon modes using magnetostriction effect, so the MMIT/A signal size scales with the microwave powers of both RF source and the VNA, capable of producing signals with high signal-to-noise ratio (SNR), but yields no information about thermal phonons. It is a valuable tool for tuning the system and identifying the mechanics.

The second scheme is the homodyne measurement.\cite{Potts2021} The same single frequency microwave tone at $\omega_\textrm{d}$ is injected into the cavity, whose photons have a chance of interacting with thermal phonons to produce sidebands at $\omega_\textrm{d} \pm \Omega_\textrm{b}$. These sidebands are then frequency down-shifted by $\omega_\textrm{d}$ using a mixer, with only the mechanical frequency remaining in the signal, which is then recorded. Signals measured with the homodyne scheme carry information of phase and amplitude fluctuations, and represent thermomechanical noise.

In our setup, the homodyne measurements of the mechanics were performed using a Zurich lock-in amplifier (LIA) and an IQ mixer (see Fig.~\ref{fig:setup}). First, we swept the photon-magnon hybrid spectrum and tuned the bias magnetic field to make the polariton separation close to the mechanical frequency $\Om b$ to prepare for the triple resonance condition. The frequency of the microwave source (RF) was then set to equal that of either upper or lower polariton peak. The in-phase (I) and out-of-phase (Q) channels of the mixer output were sent to input channels of the LIA, and the phase of the local oscillator (LO) relative to that of the RF was tuned such that almost all of the DC component appeared in the I-quadrature, while the Q-quadrature has only minimal DC. Such tuning ensures that Q carries all the mechanical signals while I carries none. The Q component is then demodulated by the LIA and the down-mixed frequency is recorded. Due to heating of the YIG sphere, the drive power had to be set to low values ($\sim\qty{-10}{dBm}$ at the cavity) in order to keep the mechanical mode thermalized to its environment. This low drive power, in conjunction with various electronic and thermal noises, resulted in a relatively SNR, necessitating the averaging of a large amount of data to resolve the signal. Roughly $50,000$ averages over the time span of a week were needed to achieve an SNR of about $10$ when the drive power was $\qty{-10}{dBm}$ at the device.

YIG has been known to have its magnon linewidth peak at temperatures around $\qty{40}{K}$,\cite{Boventer} a consequence of how rare-earth ions relax in iron garnets\cite{Seiden1964,Klingler2017}. This temperature dependence makes it possible to use the magnon linewidth as a secondary thermometer, which is particularly useful when the thermalization of the YIG sphere is in question.  For example, there could exist a large temperature difference between the temperature of the sphere and that of its surrounding as read by an external thermometer, whereas a temperature reading derived from the sphere's magnon linewidth is a direct result of its own internal temperature.

The $S_{11}$ spectrum can be measured by a VNA sweep around the resonance frequency, which can then be fit with Eq.~\ref{eq:spectrum} to obtain the parameters $\eta_\textrm{mis}$, $\omega_\textrm{a,m}$, $\kappa_\textrm{int,ext}$, $g_\textrm{am}$, and $\gamma_\textrm{m}$. An example of the spectrum is shown in Fig.~\ref{fig:cavity}d. The $\gamma_\textrm{m}$  obtained in this way is the intrinsic magnon linewidth and differs from the linewidth of either polariton peaks as shown in Fig.~\ref{fig:cavity}d. We also note that all  coupling and loss rates used are full-width values; some works use the half-width values, thus differing by a factor of two. The magnon linewidth showed a clear dependence on the magnon-photon detuning $\Delta \equiv \omega_\textrm{m} - \omega_\textrm{a}$, which can be controlled by changing bias magnetic field $B$, which in turn changes $\omega_\textrm{m}$. This apparent dependence on $\Delta$ is spurious; it comes from our fitting algorithm consistently biasing its results under certain conditions (see Supplementary Information for details and mitigation).

As a calibration process for the temperature dependence of $\gamma_\textrm{m}$, we cooled down a YIG sphere to between $\sim \qty4K$ and $\qty{25}K$. The temperatures are stabilized at various values using a PID controlled heater, and at each temperature value, multiple $S_{11}$ spectra were taken to cover the range of magnon-photon detuning $\Delta$ from $\sim\qty{-25}{MHz}\times2\pi$ to $\sim\qty{+25}{MHz}\times2\pi$. The temperature for these calibration points were read by a Cernox type thermometer produced by Lakeshore and further calibrated at National Research Council Canada (NRC) (the ``NRC thermometer'') on the International Temperature Scale of 1990 \cite{HP_Thomas_1990,HP_Thomas_1990_erratum} at the triple points of hydrogen (\qty{13.8033}K) and neon (\qty{24.5561}K); and at \qty{10}K, \qty{17}K, and \qty{20.3}K by comparison with a wire scale standard traceable to the NRC interpolating constant-volume gas thermometer.\cite{KD_Hill}

\begin{figure}[tbh]
\begin{center}
\includegraphics[width=\columnwidth]{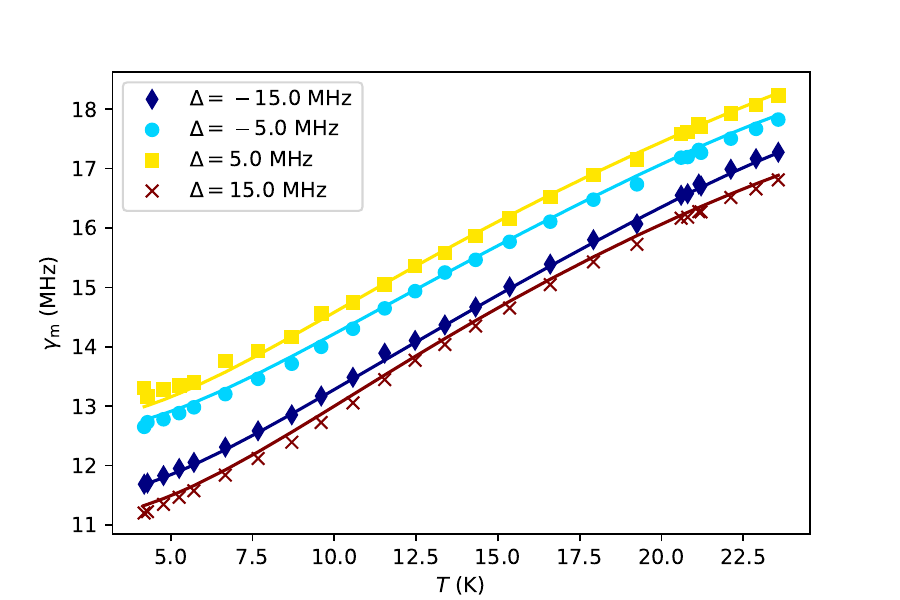}
\caption{\label{fig:gamma_m_data}A subset of measured magnon linewidth $\gamma_\textrm{m}$ at various temperature $T$ and photon-magnon detuning $\Delta$. The markers represent the measured data points, while the solid lines show fifth-order polynomial fitting for those points in the same color.}
\end{center}
\end{figure}

The above procedure resulted in a collection of $\gamma_\textrm{m}$ as a function of temperature and detuning $\Delta$, a sample of which is shown in Fig.~\ref{fig:gamma_m_data}. The value of $\gamma_\textrm{m}$ depends on both $T$ and $\Delta$, with $T$ having the stronger influence between the two, for the range we have measured. It should be noted that $\gamma_\textrm{m}$ would reduce further at lower temperatures,\cite{Boventer} well below our $\Om b \approx \qty{12.569}{MHz}$ (as seen in Fig.~\ref{fig:homodyne_result}), while at $\sim\qty{20}K$ $\gamma_\textrm{m}$ is well above $\Om b$. This means that as temperature drops from $\sim\qty{40}K$ to well below \qty4K, our setup will transition from sideband-unresolved regime into sideband-resolved, yielding interesting implications.

\begin{figure}[tbh]
\begin{center}
\includegraphics[width=\columnwidth]{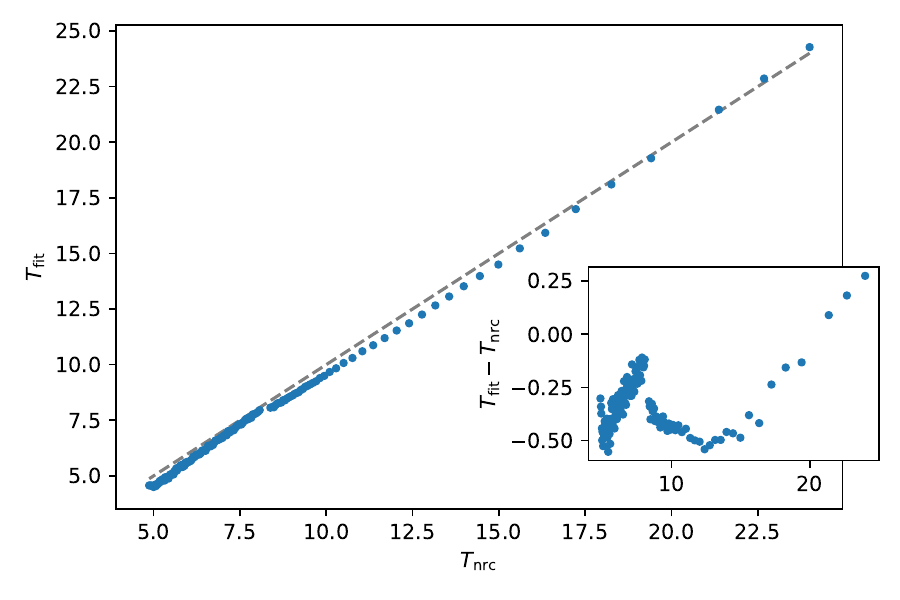}
\caption{\label{fig:Tfit}Temperature obtained by fitting of magnon linewidth compared to temperature read by the NRC thermometer. Blue dots are temperature obtained by fitting $\gammam$ and $\Delta$, while dashed line is a guide for the eye showing where the actually temperature should be. The inset shows the difference between the fit temperature and the thermometer temperature, exhibiting a discrepancy of less than $\qty{0.5}K$. There is a noticeable discontinuity of slope at $\sim\qty{8.5}K$; the bias current had to be reduced significantly at that point during the cooldown so as to not impede the superconductive transition of the bias magnet wire. This greatly changed the magnon frequency and therefore the detuning $\Delta$, which consequently showed up as a slope change of the $T_\textrm{fit}$ due to fitting artifact.}
\end{center}
\end{figure}

We devised a two-step polynomial fitting procedure to derive $T$ from any reflection spectrum taken later (see Supplementary Information for details). To validate the the fitting procedure, we performed a test on our magnon linewidth thermometer (see Fig.~\ref{fig:Tfit}). We warmed up the cryostat to above $\qty{25}K$ and let it cool back down to $\sim \qty4K$, while periodically measuring the $S_{11}$ spectra. With each spectrum taken, the temperature of the surrounding was also measured using the NRC thermometer (recorded as `$T_\textrm{nrc}$'). The spectra were later fit against Eq.~\ref{eq:spectrum} to obtain $\gammam$ and $\Delta$ values, which then yields a temperature $T_\textrm{fit}$. As can be seen in Fig.~\ref{fig:Tfit}, the fitting procedure gives a temperature value in good agreement with the NRC thermometer reading. The VNA output power at the device during all measurements was set to $\sim\qty{-47}{dBm}$ to ensure the VNA was not heating up the sphere by itself. We verified it by measuring $\gamma_\textrm{m}$ at different VNA powers and saw no variation of the linewidth, indicating no heating from the VNA. While not a high precision thermometer, the magnon linewidth clearly works well enough for our purpose, i.e.,~as a secondary thermometer that can determine if the YIG sphere is sufficiently thermalized with its environment.

\begin{figure}[tbh]
\includegraphics[width=0.9\columnwidth]{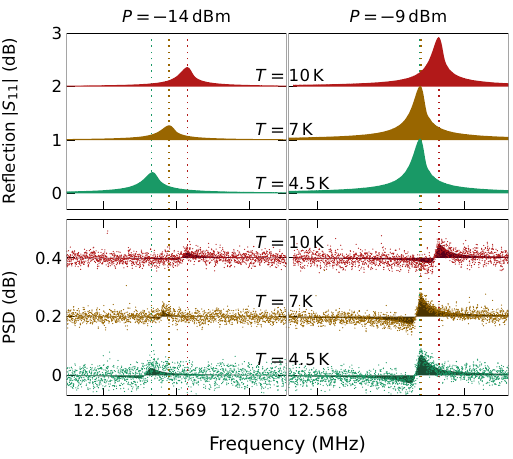}
\caption{\label{fig:homodyne_result}Measured magnomechanical signals. The top panels are the normalized $S_{11}$ scattering coefficient measured through MMIT/A on the VNA, showing the mechanical peaks with their polariton background removed, while the bottom two panels are the thermomechanical power spectral densities (PSD) measured using the homodyne setup. The dots in the PSD plots are the raw data taken, while the shaded areas are fittings of the data against a Loretzian function with Fano-like distortions. The dotted
vertical lines are guides to the eye, showing the locations of the peaks.
The temperatures in the legend correspond to the temperature of the cryostat. The plot for each temperature has been vertically shifted by a different amount to avoid clutter. The MMIT spectra have excellent signal-to-noise because they are driven through microwave interference \cite{Zhang2014,Potts2021}, but the homodyne measures the undriven, thermomechanical signals that are difficult to resolve.  As the probe power is raised these become easier to resolve, but with the consequence of heating the YIG sphere. Note that for the $P=-\qty{9}{dBm}$ case, the mechanical
signals are almost identical at $\qty{4.5}K$ and $\qty{7}K$; this is because the YIG sphere was 
heated by the RF drive, and its temperature was higher than it surrounding as read by the
thermometer due to insufficient thermalization (see Table~\ref{tab:compare_temperature}).
}
\end{figure}

We measured undriven, thermomechanical magnomechanics at temperatures of $\sim\qty4K$, $\qty7K$, and $\qty{10}K$ (as read by the NRC thermometer; for comparison against temperatures derived from magnon linewidths, see Table~\ref{tab:compare_temperature}), with two RF drive powers at the cavity, $\Pd$, at $\qty{-14}{dBm}$ and $\sim\qty{-9}{dBm}$. For each temperature/power combination, we first stabilized the stage temperature using a PID controlled heater. Once the temperature was stable,  data acquisition would begin, recording the power spectral density (PSD) of the Q output of the IQ mixer as measured by the LIA, as shown in Fig.~\ref{fig:setup}.  Also periodically taken were the VNA measurements of the polariton spectra and the MMIT spectra; the former would be later used to obtain the magnon loss rate and therefore the magnon temperature, while the latter was useful to locate the frequency of the mechanical modes so the homodyne measurement could target the correct frequency range (see Supplementary Information). The VNA output is turned off when not taking these measurements to avoid unwanted beat signals produced by the VNA probe frequencies with the drive signal.

The polariton mode was tuned such that the two peaks were separated by approximately the mechanical resonant frequency $\Om b$, and the drive tone is set to coincide with the lower peak, thus the blue sideband of the drive is at the upper polariton peak, meeting the triple resonance condition. Even so, the homodyne signal was small compared to the noise present, requiring averaging to resolve. For comparison, similar measurements done by our group before at room temperature had much clearer homodyne signals.\cite{Potts2021} This is because the current experiment has stronger clamping of the mechanical modes due to the method of anchoring of the YIG sphere (the cited experiment encased the sphere in a glass tube to minimize clamping), and much lower drive power to reduce heating.

Despite precautions taken to minimize heating of the YIG sphere by the RF drive, we still observe noticeable temperature rise when the drive signal was turned on (see Table~\ref{tab:compare_temperature}). The frequency shift of the mechanical spectrum indicates if heating is occurring (Fig.~\ref{fig:homodyne_result}), although the mechanism of this shift is currently unclear.  As one would expect, the higher input microwave power caused more heating in the sphere, producing up to $\sim\qty{10}{K}$ temperature difference between the YIG sphere and the stage temperature. However, higher drive power also yielded larger mechanical signals, as evidenced by the plots in Fig.~\ref{fig:homodyne_result}: the peaks are barely noticeable at $\qty{-14}{dBm}$ from the homodyne measurement, but are clearly visible at $\qty{-9}{dBm}$.

\begin{table}[htb]
\begin{tabular}{c@{\hskip10pt}c@{\hskip10pt}c}
$\Pd$ (dBm) & $T_\textrm{stage}$ (K) & $T_{\gammam}$ (K) \\\hline
      & $4.5$ & $9.4$  \\
$-14$ & $7$   & $12.8$ \\
      & $10$  & $13.2$ \\\hline
      & $4.5$ & $15.6$\\
$-9$  & $7$   & $15.9$ \\
      & $10$  & $18.5$ \\
\end{tabular}
\caption{\label{tab:compare_temperature}Comparison between temperatures read
by the stage thermometer and that derived from the magnon linewidth.}
\end{table}

The homodyne data in Fig.~\ref{fig:homodyne_result} show resonances at the same frequencies as measured by the higher signal-to-noise MMIT, confirming that we have resolved the thermomechanical signals at cryogenic temperatures. However, due to SNR concerns, the current experiment used drive powers from \num{-14} to \qty{-9}{dBm}, resulting in noticeable heating of the sphere, while lower input would have caused impractically long measurements with our current setup. The SNR should be improved in future optimizations, such as better magnon-photon and external couplings, and less clamping in the mounting of the YIG sphere and therefore higher quality factors in mechanical modes. It will also be important to improve thermalization of the YIG sphere in future studies so that it can be cooled to even lower temperatures. Judging by Table~\ref{tab:compare_temperature}, at $\sim\qty{4.5}K$, lowering the drive power by \qty5{dB} reduced the temperature rise due to heating by about half (from $\qty{11}K$ to $\qty{4.9}K$); we are currently exploring new methods of anchoring the YIG sphere, aimed at two goals: better thermalization, and the elimination of mechanical mode clamping (i.e.,\ not using adhesives). It is expected that the former would improve heat dissipation by orders of magnitude, while the latter would reduce the mechanical linewidth from kHz range to below \qty{100}{Hz}. With the planned scheme, we expect to be able to cool the sphere down to well below \qty1K with much smaller temperature rise, which in turn will introduce further benefit of smaller magnon linewidth.\cite{Tabuchi2014} Even if we can use these techniques to thermalize to the \qty{10}{mK} stage of the dilution refrigerator on which this experiment is already mounted, and measure without heating, we would achieve a phonon occupation of approximately~\num{17}.  To cool to the ground-state with such a low-frequency resonator would therefore require active techniques such as feedback cooling, or sideband-resolved red-detuned cooling, as have been employed in other cavity optomechanical systems\cite{teufel2011sideband,chan2011laser,kim2017magnetic}.  As such, it is likely that an entirely different architecture will be needed to achieve the quantum regime for cavity magnomechanics, one with higher frequency mechanics, better magnomechanical coupling, and better opportunities for thermalization.  One possibility is through nano-fabricated structures such as those presented recently.\cite{rashedi2024}

Despite the plethora of theoretical studies of magnomechanical interactions, there have been relatively few experimental observations of it, and none at cryogenic temperatures. Here, we have demonstrated observation of  magnomechanical signals in a $\qty{250}{\um}$-diameter YIG sphere at temperatures approaching liquid helium temperature, measuring both driven MMIT signals and thermomechanical noise spectra. We have also studied the magnon dissipation rate as a function of temperature, and have used it as a secondary -- but intrinsic -- thermometer to infer the phonon temperatures. Improvements to thermalization of the YIG sphere will enable the use of magnomechanics under cryogenic conditions for its quantum technologies, such as in quantum memory \cite{Groblacher_quantum_memory,LeBlanc} and noise correlation thermometry \cite{Potts2020}, among others.

\section*{Supplementary Material}
The supplementary material contains detailed discussions about the magnon linewidth's apparent dependence on magnon-photon detuning, and our method of fitting the magnon linewidth to obtain the phonon temperature.

\bigskip

We gratefully acknowledge financial support from the Natural Sciences and Engineering Research Council of Canada (NSERC), including grants ALLRP 558609–21, ALLRP 592535-2023, RGPIN-2022-40378, CREATE-495446-17, and a Quantum Consortium project (CanQuEST).  Additional support was obtained from the Alberta Innovates ADVANCE Program (212200780) and the National Research Council Canada Quantum Sensing Program (QSP-017-1).
 
\bibliography{references}

\end{document}


\def\om#1{\omega_\textrm{#1}}
\def\Om#1{\Omega_\textrm{#1}}

\title{Supplementary Information for
\\Cryogenic Magnomechanics for Thermometry Applications
\\at Low Temperatures}
\author{}
\maketitle

\section{Mapping magnon linewidth to temperature}

\noindent We observed that the magnon dissipation rate has a clear dependence on both the temperature and the magnon-photon detuning $\Delta \equiv \omega_\textrm{m} - \omega_\textrm{a}$, where $\omega_\textrm{m}$ and $\omega_\textrm{a}$ are the magnon and photon frequency, respectively. As the first step of our calibration process, we measured a large number of $S_{11}$ spectra, with temperature $T$ ranging from $\sim \qty4K$ to $\sim \qty{22.5}K$, and $\Delta$ from \qty{-15}{MHz} to \qty{15}{MHz}. Each such spectrum was then fit against against Eq.~1 in the main text to obtain its linewidth $\gamma_\textrm{m}$, among other parameters. A select subset of the $\gamma_\textrm{m}$ dataset is shown in Fig.~3 in the main text.

\begin{figure}
\begin{center}
\includegraphics[scale=.75]{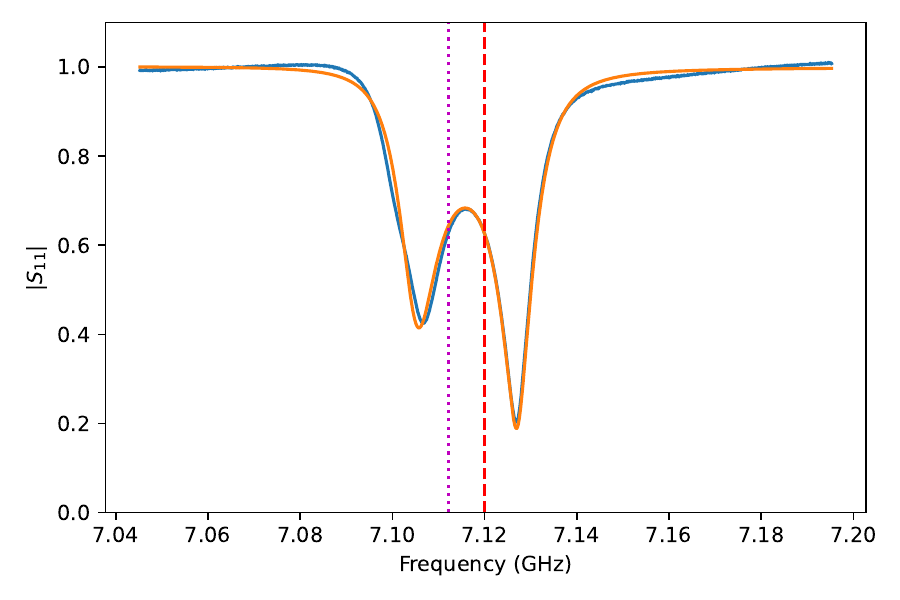}
\end{center}
\caption{\label{fig:S11}Example fitting result (red) vs measured polariton reflection spectrum (blue). It can be seen that there is a background slope on top of the polariton spectrum. This slope is part of a continuous frequency-dependent background spectrum, a result of RF signals reflecting inside coax cables and at cable junctions, and frequency dependence of various RF components. This background is difficult to physically remove, and causes a small but non-negligible discrepancy between measured spectrum and its fit result.}
\end{figure}

The magnon linewidth's dependence on temperature is expected and has be studied previously \cite{Boventer,Klingler2017,Seiden1964}. On the other hand, the apparent dependence on $\Delta$ warrants some explanation. At a given temperature, the intrinsic magnon dissipation rate should be the same regardless of magnon frequency detuning. However, the polariton spectra we measured sat on frequency-dependent background, making the $S_{11}$ spectra somewhat Fano-like, affecting the accuracy of the fitting. Moreover, at different detunings the polariton peaks are at different locations, thus are affected by different parts of this background. An example of such behavior is shown in Fig.~\ref{fig:S11}.

We have notice that, at any given temperature, the fitting inaccuracy caused by this background is consistently biased at each detuning value, meaning that the $\gamma_\textrm{m}$ value from fitting has a deviation from the true value, which is unknown but reproducible at each detuning. Because our goal with $\gamma_\textrm{m}$ is to derive temperature, some consistent bias of the $\gamma_\textrm{m}$ value is acceptable, as long as we can map it (in conjunction with detuning) back to a unique temperature value with reasonable accuracy.

To this end, we fit the calibration data $T$ as a two-step polynomial function of $\gamma_\textrm{m}$:
\begin{equation}\label{eq:Tfit}
T = \sum_{k=0}^5 a_k \gamma_\textrm{m}^k,
\end{equation}
where the coefficients $a_k$ is itself a polynomial fit over $\Delta$
\begin{equation}
a_k = \sum_{\ell=0}^5 c_{k\ell}\Delta^\ell.
\end{equation}
A polynomial of order five for both fittings was chosen empirically to have the lowest order to give satisfactory fitting of the calibration data. The efficacy of this method is shown in Fig.~4 in the main text.

\section{Measuring mechanics using VNA}
\begin{figure}[h]
\includegraphics[scale=1]{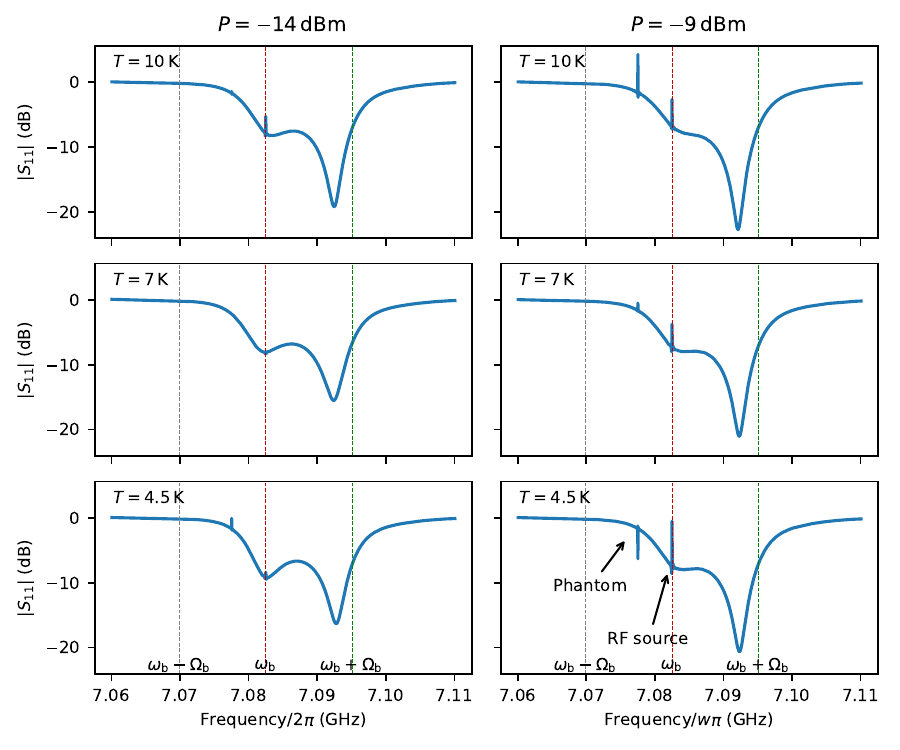}
\caption{%
    \label{fig:SI_polaritons}%
    Polariton modes, on top of which the mechanical signals were measured. The temperatures were read from the thermometers, while the actual temperatures of the phonons were higher due to the heating by the RF drive (see previous section and the main text). The spectra of the mechanical signal in Fig.~\ref{fig:SI_mech} were taken at the green lines in these plots. Here one can see another apparent RF line at exactly \qty5{MHz} below the $\om d$. This is a result of the VNA's internal circuits misinterpreting the RF drive that did not originate from its own microwave source. This phantom peak does not manifest itself when viewed on a spectral analyzer. Note that the MMIT/A signals are in these spectra, but the VNA scan setting used here did not have the frequency resolution to resolve them.
}
\end{figure}

The Magnomechanically induced transparency/absorption (MMIT/A) signals can be measured with a vector network analyzer (VNA). When a tone with a single frequency $\om d$ is injected into the cavity, its photons interact with the resonant phonon mode of the YIG sphere at frequency $\Om b$ and produce sidebands at $\om d \pm \Om b$ (Fig.~\ref{fig:SI_polaritons}. For the arrangements shown in Fig.~\ref{fig:SI_polaritons}, the drive $\om d$ is at the peak of one polariton mode, while the $\om d + \Om b$ sideband (marked by the green lines) is near the other peak, thus receives significant cavity enhancement. The other sideband at $\om d - \Om b$ receives no such enhancement, and is much smaller in amplitude.

In a true triply resonant condition, one of the sidebands (the green line in our case) should also coincide with a polariton peak. To achieve it, we should have detuned magnon frequency $\om m$ further away from cavity resonance frequency $\om a$, increasing the separation between the polariton peaks such that it equals $\Om b$. However, we found that when we tuned the system so, the magnon-like peak became too small, and the system became unstable when the RF drive was turned on. As a compromise, we settled for the configurations show in Fig.~\ref{fig:SI_polaritons}, where the the $\om d+\Om b$ location is not at of the upper polariton peak, but close enough to receive substantial enhancement.

\begin{figure}[h]
\includegraphics[scale=1]{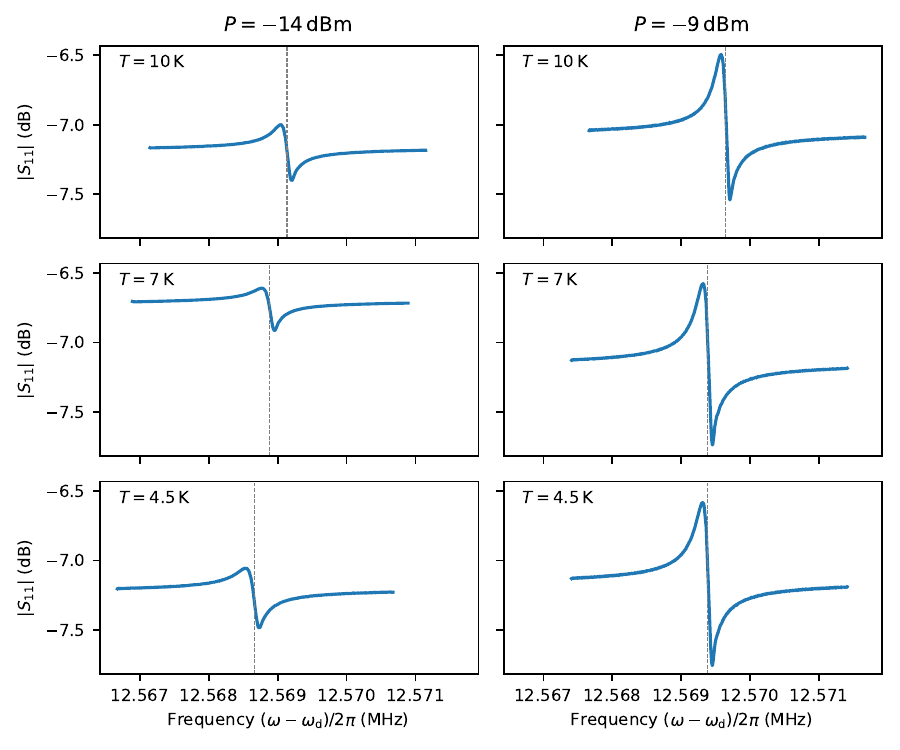}
\caption{%
    \label{fig:SI_mech}%
    Mechanics in the form of MMIT/A signals, recorded by the VNA. Because these peaks were on top of the slope of the polariton modes, they all have Fano-like shapes. For the plots of MMIT/A signals with the background slope removed, see Fig.~5 in the main text. Here the gray vertical lines show where the peaks were after removing the background slope.
}
\end{figure}

The frequency resolution in Fig.~\ref{fig:SI_polaritons} is too coarse to show the mechanical signals themselves. For that, we can use VNA to scan a much narrower range around the green lines, the results of which are shown in Fig.~\ref{fig:SI_mech}. It is clear that as the temperature increases, the mechanical frequency shows a tendency of increasing slightly. It is currently unclear why this happens, and our current experimental setup is not sufficiently equipped to properly investigate it. However, we do have some indications of unexpected mechanical complexities, which can be seen in the 2D plots of the complex signal of the $S_{11}$ spectra. 

\begin{figure}[h]
\includegraphics{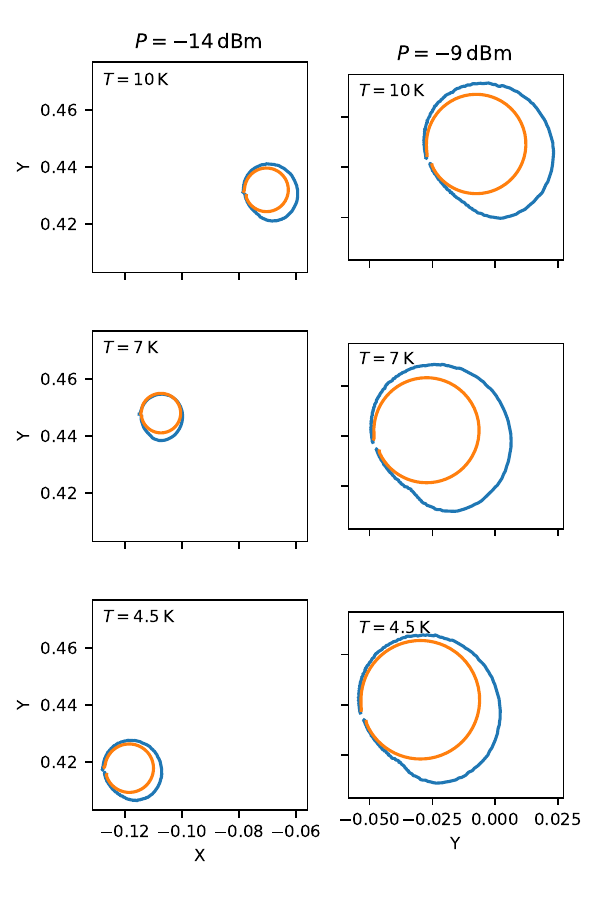}
\caption{%
    \label{fig:SI_xy}%
    2D plots of the complex MMIT/A signals as measured by the VNA. Blue are the actual data points, while orange are rough attempts of fitting them against a Lorentzian function. 
}
\end{figure}

When an ideal MMIT/A signal is plotted in the complex plane, its Lorentzian function form should produce a perfect circle. When our recordings were so plotted (Fig.\ref{fig:SI_xy}), they showed clear deviations. The shape of these $xy$-plots suggest very closely spaced degenerate mechanical modes, caused either by clamping of our mounting method, or the non-uniformity of the YIG sphere itself. This complex mechanical mode may also be related to the mechanical frequency shifting with temperature. However, the question can only be answered by future studies.
\bibliography{ref_SI}